\documentclass[iop,english,twocolappendix,numberedappendix,showpacs,superscriptaddress,appendixfloats,tighten,apj,twocolumn]{aastex63}
\usepackage{amsmath,amsfonts,amssymb,xcolor,soul}
\usepackage{dcolumn,ulem}
\usepackage{bm}
\usepackage{float}
\usepackage{hyperref}
\usepackage{subfigure}
\def\ADD#1{{\textcolor{black}{#1}}}

\newcommand{\p} {\partial}
\newcommand{\el}{\boldsymbol{\ell}}
\newcommand{\nab}{\boldsymbol\nabla}
\newcommand{\elb}{\boldsymbol\ell}
\def\vv{{\bf v}}

\def\bb{{\bf v_A}}
\def\rr{{\bf \el}}
\def\div{{\pmbmath{\nabla} \cdot}}
\def\divprim{{\pmbmath{\nabla}' \cdot}}
\def\divl{{\pmbmath{\nabla}_\rr \cdot}}
\def\grad{{\pmbmath{\nabla}}}

\def\be{\begin{equation}}
\def\ee{\end{equation}}
\def\ba{\begin{eqnarray}}
\def\ea{\end{eqnarray}}

\def \pmbmath{\mathpalette\pmbmathaux}
\def \pmbmathaux#1#2{
         \pmbtext{$#1#2$}}
\def \pmbtext#1{\leavevmode
     \setbox0\hbox{#1}
     \kern0,4pt \copy0 \kern-\wd0
     \kern-0,2pt \raise0,3pt \box0 }
     
\begin{document}
\title{General exact law of compressible isentropic magnetohydrodynamic flows: theory and spacecraft observations in the solar wind}

\author{P. Simon} 
\email{pauline.simon@lpp.polytechnique.fr}
\affiliation{Laboratoire de Physique des Plasmas, CNRS, \'Ecole polytechnique, Universit\'e Paris-Saclay, Sorbonne Universit\'e, Observatoire de Paris-Meudon, F-91128 Palaiseau Cedex, France}

\author{F. Sahraoui} 
\affiliation{Laboratoire de Physique des Plasmas, CNRS, \'Ecole polytechnique, Universit\'e Paris-Saclay, Sorbonne Universit\'e, Observatoire de Paris-Meudon, F-91128 Palaiseau Cedex, France}

\date{\today}

\begin{abstract}
Various forms of exact laws governing magnetohydrodynamic (MHD) turbulence have been derived either in the incompressibility limit, or for  isothermal compressible flows. Here we propose a more general method that allows us to obtain such laws for any turbulent isentropic flow (i.e., constant entropy). We demonstrate that the known MHD exact laws (incompressible and isothermal) and the new (polytropic) one can be obtained as specific cases of the general law when the corresponding closure equation is stated. We also recover all known exact laws of hydrodynamic (HD) turbulence (incompressible, isothermal and polytropic) from this law in the limit ${\bf B}=0$. We furthermore show that the difference between the two forms (isothermal and polytropic) of the MHD exact laws of interest in this work resides in some of the source terms and in the explicit form of the flux term that depends on internal energy. Finally, we apply these two forms to Parker Solar Probe (PSP) data taken in the inner heliosphere to highlight how the different closure equations affect the energy cascade rate estimates.
\end{abstract}


\section{Introduction}

The formalism of exact law describing turbulent flows was first developed in the context of neutral fluid dynamics : incompressible HD \citep{kolmogorov_dissipation_1991,frisch_turbulence_1995,antonia_analogy_1997,galtier_introduction_2016}, isothermal \citep{galtier_exact_2011} and polytropic \citep{banerjee_kolmogorov-like_2014} compressible HD. Then it has been extended to magnetized plasmas described within various approximations: incompressible MHD (IMHD) \citep{politano_von_1998,yoshimatsu_examination_2012,banerjee_alternative_2017}, incompressible Hall-MHD (IHMHD)\citep{galtier_von_2008,hellinger_von_2018,ferrand_exact_2019,banerjee_alternative_2017}, incompressible \citep{andres_exact_2016,andres_von_2016} and compressible \citep{banerjee_scale--scale_2020} two fluids, isothermal compressible MHD (CMHD) \citep{banerjee_exact_2013,andres_energy_2018,andres_alternative_2017}, and isothermal compressible Hall-MHD (CHMHD) \citep{andres_exact_2018}. The main driver of such efforts, other than their intrinsic theoretical interest, is to better model the turbulent heating of the solar wind and magnetospheric plasmas \citep{richardson_radial_1995,bruno_solar_2013,sahraoui_magnetohydrodynamic_2020}. Indeed, the formalism of exact law allows us to derive (within the classical assumptions that will be recalled further below) exact relations that couple in a full nonlinear way, the turbulent fields to the energy transfer rate between scales. This cascade rate is assumed to be equal to the rate by which energy is injected into the system (at the largest scales) and to the rate by which it is dissipated  (at the smallest scales). Although these fluid models have strong limitations, in particular when applied to collisionless space plasmas, they nevertheless provide useful means by which one can apprehend the challenging problem of quantifying energy dissipation and particle heating in spacecraft observations. This is witnessed by the significant progress that has been achieved in recent years on these questions, from the evidence of turbulence cascade in space plasmas \citep{smith_dependence_2006,sorriso-valvo_observation_2007,sahraoui_diagnosis_2008,stawarz_turbulent_2010,marino_occurrence_2012,coburn_third-moment_2015}, to the study of how density fluctuations enhance the dissipation rate in the solar wind and planetary magnetospheres \citep{carbone_scaling_2009,banerjee_scaling_2016,hadid_energy_2017,hadid_compressible_2018,andres_solar_2020}, and the first estimation of the turbulent dissipation rate at kinetic (sub-ion) scales in the Earth's magnetosheath turbulence \citep{andres_energy_2019,sorriso-valvo_turbulence-driven_2019,bandyopadhyay_situ_2020,quijia_comparing_2021} \ADD{and other key region of the magnetosphere \citep{sorriso-valvo_turbulence-driven_2019,quijia_comparing_2021}}.

Here, we extend the previous efforts by proposing a general method that allows us to obtain the exact law for any isentropic flow using the internal energy equation, but without a prior specification of the state equation (isothermal or polytropic). Among other results, this formulation provides a new exact law for compressible polytropic MHD flows, which allows us to relax the isothermal closure when applying the exact law formalism to space plasma observations \citep{hadid_energy_2017,hadid_compressible_2018,andres_solar_2020}. \ADD{We also show a first application of the new law to Parker Solar Probe (PSP) to highlight similarities and differences with the isothermal closure.}

\section{Generalized exact law for isentropic flows}\label{theory}

Our analysis is based on the following compressible MHD equations 
\ba
\p_t \rho + \div (\rho \vv) &=&0 \, , \label{mhd1} \\
\partial_t (\rho \vv )+ \div (\rho \vv \vv) &=& \div (\rho \bb \bb) - \grad (P + P_M) \nonumber \\ &&+ {\bf d_k} + {\bf f} \, , \label{mhd2} \\
\div (\rho \bb) &=& - \rho \div \bb \, , \label{mhd3} \\
\partial_t (\rho \bb) + \div (\rho \vv \bb) & = &  \div (\rho \bb \vv) + \rho \vv \div \bb  \nonumber \\ &&- \frac{1}{2} \rho \bb \div \vv + {\bf d_m} \, , \label{mhd4}
\ea
where $\rho$ is the mass density, $\vv$ the velocity field, $\bb$ the Alfv\'en velocity, $P$ the pressure, $P_M$ the magnetic pressure, ${\bf d_k}$ the kinetic viscous dissipation, ${\bf d_m}$ the magnetic resistive dissipation, ${\bf f}$ a stationary homogeneous external force assumed to act on large scales.

Note that the above system is not closed so long as a closure equation relating, for instance, the second order moment $P$ to the mass density $\rho$ is not specified. Instead of adopting such an equation at this stage as done in all previous derivations of exact laws for compressible HD and MHD turbulence  \citep{galtier_exact_2011,banerjee_exact_2013,banerjee_kolmogorov-like_2014,andres_alternative_2017,andres_exact_2018}, here we elect to remain more general and assume only that we are to deal with isentropic flows in which the entropy remains constant (note that sometimes these flows are called barotropic, see \cite{zakharov_kolmogorov_1992,eyink_cascades_2018}). To this purpose we introduce a general equation that governs the internal energy variations in such flows. We show below that such a law encompasses incompressible and compressible (isothermal and polytropic) closures, whose von Kármán-Howarth-Monin (KHM) equations will be obtained as simple limits of the general form that we derive in this work. To do so we need to recall some thermodynamics notions in view of highlighting some subtle differences between turbulent cascade in compressible and incompressible flows and the concept of energy dissipation within the formalism of exact laws. 

\subsection{Some remarks on the thermodynamics}\label{thermo}

Let us start with the definition of internal energy variation $\delta U$ in a given system, based on the first principle of thermodynamics
\ba
\delta U&=&\delta Q+\delta W = T\delta S - P\delta V \label{ther0}
\ea  
which reflects the two distinct contributions to the internal energy variation: the heat component $\delta Q=T\delta S$ due to entropy variation $\delta S$ and the work of the pressure force $\delta W$ ($V$ and $T$ are the volume and temperature of the considered thermodynamical system). For isentropic flows, i.e. a constant entropy, the internal energy variation reduces to the work of the pressure force:
\be
\delta U=- P\delta V \label{ther1}
\ee 
with the definition of the mass density $\rho=\frac{m}{V}$ one obtains by simple variation 
\be
\delta V=- \frac{m}{\rho^2}\delta \rho \, . \label{ther2}
\ee 
Injecting relation (\ref{ther2}) into equation (\ref{ther1}) and introducing the specific internal energy $u=U/m$ (i.e. internal energy per unit mass) yield
\be
\delta u= \frac{P}{\rho^2}\delta \rho \, . \label{ther3}
\ee 
Introducing the time and spatial derivatives one can obtain from equation (\ref{ther3}) the following relations
\ba 
\p_t u &=&\frac{P}{\rho^2}\p_t \rho \, , \label{ther4}\\
\grad u&=&\frac{P}{\rho^2}\grad \rho \, . \label{ther5}
\ea  
Combining relations (\ref{ther4})-(\ref{ther5}) with the continuity equation (\ref{mhd1}) one obtains the following ``continuity'' equation for the specific internal energy
\be
\p_t u +\vv .\grad(u)=-\frac{P}{\rho}\div \vv \, , \label{mhd6}
\ee  
which can be written for convenience as
\be
\partial_t (\rho u) + \div (\rho u \vv) =  - P \div \vv \, . \label{ther6}
\ee
Some remarks can be made here based on equation (\ref{mhd6}-\ref{ther6}). First, we recall that this equation was derived assuming a constant entropy and an isotropic pressure. This means that this equation forbids any (irreversible) dissipation that requires increase of entropy. At a first sight this observation might sound in contradiction with the idea that turbulence requires dissipation, which is reflected by the presence of the resistive and viscous terms in equations (\ref{mhd2}) and (\ref{mhd4}) \citep{eyink_cascades_2018}. The contradiction can actually be solved thanks to the assumptions used is the derivation of the turbulence exact laws (as will be recalled below): they are valid only in a range of (inertial) scales where the dissipation is negligible with respect to nonlinear transfers, the former becoming comparable or larger than the latter only at very smallest scales. In this sense, internal energy equation (\ref{mhd6}-\ref{ther6}) should be understood as valid only in the inertial range. Entropy can be produced but only at (and confined to) small scales. A more rigorous solution to this problem would \ADD{consist in introducing a cooling term (e.g., thermal radiations)} that would extract entropy production (and subsequent internal energy excess) at sufficient rate so that equation (\ref{mhd6}-\ref{ther6}) remains valid at all scales \citep{eyink_cascades_2018}. 

The second remark that can be made is that in the incompressible limit (i.e., $\grad. \vv=0$), the internal energy is fully conserved for an isentropic flow: $d_t (u)=\p_t u +\vv .\grad u =0$. This observation means that in incompressible HD flows (a generalization to MHD is straightforward) the total energy is the sum of the kinetic and internal energies, but the cascade concerns only the kinetic energy \citep{kolmogorov_dissipation_1991,frisch_turbulence_1995,politano_von_1998}. The explanation is rather simple: under the assumptions that (kinetic) energy is injected at the largest scales and dissipation limited to the smallest ones, since internal energy is conserved (i.e., no exchange with kinetic energy) then the injected energy has to cascade to small scales (due to nonlinearities) where it is eventually dissipated. This ``necessity'' of cascade in incompressible flows does not hold in compressible flows, for which the internal energy is no longer conserved and exchanges with kinetic energy via the r.h.s. term of equation (\ref{mhd6}-\ref{ther6}). In this case and under the same assumptions as above, once kinetic energy is injected at the largest scales, it can nonlinearly exchange with the internal energy (through the pressure forces) in a way that may populate the inertial range but without necessarily implying a cascade toward small scales with a constant flux. An example would be planar shocks in compressible fluids (a multi-scale problem where compression dominates and heating occurs, but the notion of cascade is irrelevant).  This idea is supported by the presence in all the existing exact laws for compressible HD, MHD and Hall-MHD flows of source terms (i.e., proportional to the divergence of fields such as $\vv$ and $\bb$) that act in the inertial range and reflect the presence of compression and dilatation of the fluid \citep{galtier_exact_2011,banerjee_exact_2013,andres_exact_2018,andres_energy_2019,ferrand_compact_2021}. Those terms can fully balance the flux terms, at least in some instances of supersonic turbulence \citep{ferrand_compressible_2020}. In such situations the existence of an energy cascade with a constant flux over scales is not guaranteed. Therefore, in compressible models of turbulence, the examination of the weight of the source with respect to the flux terms is important to assess the presence or not of energy cascade over scales as known in incompressible flows.  

A final remark about compressible turbulent flows can be made regarding the nature of the energy that cascades to small scales. In the previous works (and in the present paper) the derivation of the exact law was based on using two-point correlation functions that can be related to the total energy of the system (i.e., sum of kinetic, magnetic and internal energies) \citep{banerjee_exact_2013,andres_exact_2018,andres_energy_2019,ferrand_compressible_2020,ferrand_compact_2021}. However, it is important to note that in such formulations it is the sole ``component'' of internal energy driven by the pressure forces as defined by equations (\ref{ther1}) that is considered in the cascade process. The other ``component'' due to entropy variation ($TdS$) is assumed to be negligible in the inertial range. This observation does not contradict the fact that all the energy will be eventually converted into heat at the smallest scales, which will result into an increase of the internal energy through entropy variation $T\delta S$ according to equation (\ref{ther0}). From this viewpoint, there is no difference between incompressible and compressible turbulence: energy will eventually be dissipated into heat. The difference lies only in the nature of the energy that undergoes that conversion: in the former it is the kinetic energy (so long as internal energy remains conserved, i.e. the work of the pressure force is zero), in the latter it is the total energy (the sum of kinetic and internal energies).   

\subsection{KHM equation for compressible isentropic MHD turbulence}

To derive the KHM equation for isentropic flow we use equations (\ref{mhd1})-(\ref{mhd4}) complemented by the internal energy equation (\ref{mhd6}). We define $\el$ the spatial increment connecting two points $\textbf{x}$ and $\textbf{x}'$ as ${\bf x'} = {\bf x}+{\bf \el}$ and, for any given field $\xi$, $\xi(\textbf{x}) \equiv \xi$ and $\xi(\textbf{x}') \equiv \xi'$.
We consider the mean correlation function $R_{tot} = \langle R + R' \rangle/2$ with $\langle \rangle$ an ensemble average, $\langle R \rangle = \langle \rho \vv \cdot \vv' /2 + \rho \bb \cdot \bb' /2 +  \rho u' \rangle = \langle R_k + R_B + R_u \rangle $ a correlation function taken at the point $\textbf{x}$ and $R'$ its conjugate, i.e. the same function taken at the point $\textbf{x}'$. We note that if $\textbf{x} = \textbf{x}'$, $R_{tot} = \langle E \rangle = \langle \rho v^2 /2 + \rho v_A{}^2 /2 + \rho u \rangle $, which defines the mean total energy of the system.  

Using the property $\partial_t \langle \rangle = \langle \partial_t  \rangle$, the temporal evolution of $R_{tot}$ is given by 
\ba
\partial_t R_{tot} &=& \frac{1}{4} \langle \partial_t (\rho \vv \cdot \vv') + \partial_t (\rho' \vv' \cdot \vv ) \rangle \nonumber \\ &&+ \frac{1}{4} \langle \partial_t (\rho \bb \cdot \bb') + \partial_t (\rho' \bb' \cdot \bb ) \rangle \nonumber \\ &&+ \frac{1}{2} \langle \partial_t (\rho u' + \rho' u ) \rangle \, . \label{cor}
\ea
We are interested in deriving the explicit forms of the time evolution of the correlators $\langle R_k \rangle$, $\langle R_B \rangle$ and $\langle R_u \rangle$ involved in equation (\ref{cor}). For this purpose we use equations (\ref{mhd1})-(\ref{mhd4}) and  (\ref{mhd6}) written at the positions $\textbf{x}$ and $\textbf{x}'$. We recall that in our formalism we can write for any entity A: $\partial_x A' = \partial_{x'} A = 0$, and define $\delta A \equiv A'-A$ . Under the hypothesis of space homogeneity we also have the relations $\langle\nab'\cdot\rangle=\nab_{\elb} \cdot\langle\rangle$ and $\langle\nab\cdot\rangle=-\nab_{\elb} \cdot\langle\rangle$, where $\nab_{\elb}$ denotes the derivative operator along the increment $\elb$. Using these relations, we obtain the following expressions for the correlators involved in equation (\ref{cor}) \ADD{(see Appendix \ref{ann:details} for details)}:

\begin{widetext}
\ba
2\langle \partial_t (R_k + R'_k ) \rangle &=& \langle \partial_t (\rho \vv \cdot \vv' + \rho' \vv' \cdot \vv ) \rangle  \nonumber \\ 
    &=& \divl \langle \rho \vv \cdot \vv' \vv - \rho \vv \cdot \vv' \vv' - \rho' \vv' \cdot \vv \vv' + \rho' \vv' \cdot \vv \vv \rangle \nonumber \\
    &&- \divl \langle \rho \bb \cdot \vv' \bb - \rho \vv \cdot \bb' \bb' - \rho' \bb' \cdot \vv \bb' + \rho' \vv' \cdot \bb \bb \rangle \nonumber \\ 
    &&+ \divl \langle (P + P_M)\vv' - (P' + P'_M)\vv - \frac{\rho}{\rho'} (P' + P'_M) \vv +  \frac{\rho'}{\rho} (P + P_M) \vv' - \rho u' \vv +  \rho' u \vv' \rangle \nonumber \\
    &&+ \langle \rho' \vv' \cdot \vv \div \vv + \rho \vv \cdot \vv' \divprim \vv' - 2\rho' \vv' \cdot \bb \div \bb - 2\rho \vv \cdot \bb' \divprim \bb' - \frac{P'_M}{P'} \divprim (\rho u' \vv) - \frac{P_M}{P} \div (\rho' u \vv') \rangle \nonumber \\
     &&+ \langle {\bf d_k} \cdot \vv' + \frac{\rho}{\rho'}{\bf d'_k} \cdot \vv + {\bf d'_k} \cdot \vv +\frac{\rho'}{\rho}{\bf d_k} \cdot \vv' + {\bf f} \cdot \vv' + \frac{\rho}{\rho'}{\bf f'} \cdot \vv + {\bf f'} \cdot \vv + \frac{\rho'}{\rho}{\bf f} \cdot \vv' \rangle  , \label{eq-corK}\\
2\langle \partial_t (R_B + R'_B ) \rangle &=& \langle \partial_t (\rho \bb \cdot \bb' + \rho' \bb' \cdot \bb ) \rangle \nonumber \\ 
    &=& \divl \langle \rho \bb \cdot \bb' \vv - \rho \bb \cdot \bb' \vv' - \rho' \bb' \cdot \bb \vv' + \rho' \bb' \cdot \bb \vv \rangle \nonumber \\
    &&- \divl \langle \rho \vv \cdot \bb' \bb - \rho \bb \cdot \vv' \bb' - \rho' \vv' \cdot \bb \bb' + \rho' \bb' \cdot \vv \bb \rangle \nonumber \\ 
    &&+ \langle (\frac{1}{2} \rho' \bb' \cdot \bb - \frac{1}{2} \rho \bb \cdot \bb') \div \vv + (\frac{1}{2} \rho \bb \cdot \bb' - \frac{1}{2} \rho' \bb' \cdot \bb) \divprim \vv' \rangle \nonumber \\ 
    &&+ \langle (\rho' \vv' \cdot \bb - \rho \bb \cdot \vv') \divprim \bb' + (\rho \vv \cdot \bb'- \rho' \bb' \cdot \vv) \div \bb \rangle \nonumber \\
    &&+ \langle {\bf d_m} \cdot \bb' + \frac{\rho}{\rho'}{\bf d'_m} \cdot \bb + {\bf d'_m} \cdot \bb + \frac{\rho'}{\rho}{\bf d_m} \cdot \bb' \rangle , \label{eq-corM}\\
2\langle \partial_t (R_u + R'_u ) \rangle &=& \langle \partial_t (\rho u' + \rho' u ) \rangle  \nonumber \\ 
    &=& \divl \langle \rho u' \vv - \rho u' \vv' - \rho' u \vv' - \rho' u \vv \rangle + \langle (\rho u' - \frac{\rho}{\rho'} P') \divprim \vv' + (\rho' u - \frac{\rho'}{\rho} P) \div \vv \rangle \, . \label{eq-corU}
\ea
\end{widetext}
\ADD{The $\divl$ terms (except the third line of the equation (\ref{eq-corK}) that contains pressure terms) of the equations (\ref{eq-corK}-\ref{eq-corU}) contain a part of the developed form of the structure functions $\langle \delta (\rho \vv )\cdot \delta \vv \delta \vv \rangle$, $\langle \delta (\rho \bb ) \cdot \delta \bb \delta \vv \rangle$, $\langle \delta \rho \delta u \delta \vv \rangle$, $\langle \delta (\rho \bb ) \cdot \delta \vv \delta \bb \rangle$ and $\langle \delta (\rho \vv ) \cdot \delta \bb \delta \bb \rangle$. The other part can be written as source terms that include a dilatation factor like $\div \vv$ or $\div \bb$.
Then, introducing the structure functions in the equations (\ref{eq-corK}-\ref{eq-corU}) (see Appendix \ref{ann:details} for an example), equation (\ref{cor}) which is the sum of the equations (\ref{eq-corK}-\ref{eq-corU}) becomes:}
\begin{widetext}
\ba
4 \partial_t R_{tot}
    &=& \divl \langle \delta (\rho \vv ) \cdot \delta \vv \delta \vv + \delta (\rho \bb ) \cdot \delta \bb \delta \vv + 2 \delta \rho \delta u \delta \vv - \delta (\rho \bb ) \cdot \delta \vv \delta \bb - \delta (\rho \vv ) \cdot \delta \bb \delta \bb \rangle \nonumber \\
    &&+ \divl  \langle  (1+\frac{ \rho' }{\rho}) (P + P_M) \vv' - (1+\frac{ \rho }{\rho'})(P' + P'_M) \vv + \rho' u \vv' - \rho u' \vv \rangle \nonumber \\
    &&+ \langle (\divprim \vv')(\rho \vv \cdot \delta \vv + \rho \bb \cdot \delta \bb - \frac{1}{2} \rho' \bb' \cdot \bb - \frac{1}{2} \rho  \bb \cdot \bb' + 2\rho (\delta u - \frac{P'}{\rho'})) \rangle \nonumber \\
    &&+ \langle (\div \vv)(- \rho' \vv' \cdot \delta \vv - \rho' \bb' \cdot \delta \bb - \frac{1}{2} \rho  \bb \cdot \bb' - \frac{1}{2} \rho'  \bb' \cdot \bb - 2 \rho' (\delta u + \frac{P}{\rho}) ) \rangle \nonumber \\
    &&-  \langle (\divprim \bb')(2\rho \vv \cdot \delta \bb - \rho' \vv' \cdot \bb + \rho \bb \cdot \vv' ) - (\div \bb)(2 \rho' \vv' \cdot \delta \bb + \rho \vv \cdot \bb' - \rho' \bb' \cdot \vv ) \rangle \nonumber \\
    &&- \langle \frac{P'_M}{P'} \divprim (\rho u' \vv) + \frac{P_M}{P} \div (\rho' u \vv') \rangle + \langle {\bf f} \cdot \vv' + \frac{\rho}{\rho'}{\bf f'} \cdot \vv + {\bf f'} \cdot \vv + \frac{\rho'}{\rho} {\bf f} \cdot \vv' \rangle\nonumber \\
    &&+ \langle {\bf d_k} \cdot \vv' + \frac{\rho}{\rho'}{\bf d'_k} \cdot \vv + {\bf d'_k} \cdot \vv + \frac{\rho'}{\rho} {\bf d_k} \cdot \vv' + {\bf d_m} \cdot \bb' + \frac{\rho}{\rho'}{\bf d'_m} \cdot \bb + {\bf d'_m} \cdot \bb + \frac{\rho'}{\rho}{\bf d_m} \cdot \bb' \rangle \, . \label{KHM}
\ea
\end{widetext}

Equation (\ref{KHM}) is the KHM equation for isentropic MHD turbulence. Following the usual assumptions used in fully developed homogeneous turbulence (infinite kinetic and magnetic Reynolds numbers, a stationary state with a balance between forcing and dissipation \citep{frisch_turbulence_1995,galtier_exact_2011}), one obtains
\begin{widetext}
\ba
- 4 \varepsilon
    &=& \divl \langle \delta (\rho \vv ) \cdot \delta \vv \delta \vv + \delta (\rho \bb ) \cdot \delta \bb \delta \vv + 2 \delta \rho \delta u \delta \vv - \delta (\rho \bb ) \cdot \delta \vv \delta \bb - \delta (\rho \vv ) \cdot \delta \bb \delta \bb \rangle \nonumber \\
    &&+ \divl  \langle  (1+\frac{ \rho' }{\rho}) (P + P_M) \vv' - (1+\frac{ \rho }{\rho'})(P' + P'_M) \vv + \rho' u \vv' - \rho u' \vv \rangle \nonumber \\
    &&+ \langle (\divprim \vv')(\rho \vv \cdot \delta \vv + \rho \bb \cdot \delta \bb - \frac{1}{2} \rho' \bb' \cdot \bb - \frac{1}{2} \rho  \bb \cdot \bb' + 2\rho (\delta u - \frac{P'}{\rho'})) \rangle \nonumber \\
    &&+ \langle (\div \vv)(- \rho' \vv' \cdot \delta \vv - \rho' \bb' \cdot \delta \bb - \frac{1}{2} \rho  \bb \cdot \bb' - \frac{1}{2} \rho'  \bb' \cdot \bb - 2 \rho' (\delta u + \frac{P}{\rho}) ) \rangle \nonumber \\
    &&-  \langle (\divprim \bb')(2\rho \vv \cdot \delta \bb - \rho' \vv' \cdot \bb + \rho \bb \cdot \vv' ) - (\div \bb)(2 \rho' \vv' \cdot \delta \bb + \rho \vv \cdot \bb' - \rho' \bb' \cdot \vv ) \rangle \nonumber \\
    &&- \langle \frac{P'_M}{P'} \divprim (\rho u' \vv) + \frac{P_M}{P} \div (\rho' u \vv') \rangle \, , \label{exactlaw}
\ea
\end{widetext}
where we used the assumption $\langle {\bf f} \cdot \vv' + \frac{\rho}{\rho'}{\bf f'} \cdot \vv + {\bf f'} \cdot \vv + \frac{\rho'}{\rho} {\bf f} \cdot \vv' \rangle \simeq 4 \varepsilon$, $\varepsilon$ being the mean rate of energy injection by unit mass. Equation (\ref{exactlaw}) is the first main result of this paper : a general exact law for compressible isentropic MHD flows. It is valid in the inertial range where the forcing and dissipation are assumed to be negligible with respect to the nonlinear terms. Note that this equation depends on the plasma pressure $P$ whose explicit dependence on mass density $\rho$ is not yet stated. We show now that by specifying such a relation (i.e. choosing a closure equation) we can derive the exact laws for isothermal and polytropic MHD (and HD in the limit ${\bf B}=0$) turbulence, along with their incompressible limits.

\subsubsection{Incompressible MHD turbulence}
The incompressible MHD exact law \citep{politano_von_1998} can readily be obtained from equation (\ref{exactlaw}) in the limit of constant mass density $\rho = \rho_0$, which implies that all the source terms (i.e., those proportional to field divergence) tend to zero. Equation (\ref{exactlaw}) reduces then to  
\be
- 4 \frac{\varepsilon}{\rho_0} = \divl \langle (\delta \vv \cdot \delta \vv+\delta \bb \cdot \delta \bb) \delta \vv - 2 \delta \bb \cdot \delta \vv \delta \bb \rangle \, .\label{imhd}
\ee

\subsubsection{Compressible isothermal MHD turbulence}

In the isothermal case, the state equation is $P = c_s{}^2 \rho$ with $c_s$ the constant sound speed. Then one can readily demonstrate the  equality: 
\begin{widetext}
\ba
\divl\langle(1+\frac{ \rho' }{\rho})(P&+&P_M) \vv' - (1+\frac{ \rho }{\rho'})(P' + P'_M) \vv + \rho' u \vv' - \rho u' \vv \rangle - \langle 2\rho \frac{P'}{\rho'} \divprim \vv' \rangle - \langle 2 \rho' \frac{P}{\rho} \div \vv \rangle \nonumber \\
&=&\langle (P_M - P) \divprim \vv' \rangle + \langle (P'_M - P') \div \vv \rangle + \langle (\frac{\bb^2}{2} + u) \divprim (\rho' \vv') \rangle + \langle (\frac{\bb'^2}{2} + u') \div (\rho \vv) \rangle \, .
\ea
\end{widetext}

With this source terms modifications and introducing the notation $\beta = P/P_M$, $\bar \rho = (\rho'+\rho)/2$ for the average density, $H = \rho \vv \cdot \bb$ the cross helicity, $R_E = \rho \vv \cdot \vv' /2 + \rho \bb \cdot \bb' /2 +  \rho u'$, $R_B = \rho \bb \cdot \bb' /2 $ and $R_H = \rho \vv \cdot \bb' /2 + \rho \bb \cdot \vv' /2 $, the non-averaged correlators for the total energy, magnetic energy and cross helicity, we recognize the isothermal exact law derived by \cite{andres_alternative_2017}: 
\begin{widetext}
\ba
- 2 \varepsilon
    &=& \frac{1}{2}\divl \langle \big[\delta (\rho \vv ) \cdot \delta \vv  + \delta (\rho \bb ) \cdot \delta \bb  + 2 \delta \rho \delta u \big]\delta \vv - \big[\delta (\rho \bb ) \cdot \delta \vv  + \delta (\rho \vv ) \cdot \delta \bb \big]\delta \bb \rangle \nonumber \\
    &&+ \langle (\divprim \vv')\big[R_E -\frac{R_B + R'_B}{2}- E + \frac{P_M - P}{2}\big] \rangle + \langle (\div \vv)\big[R'_E -\frac{R'_B + R_B}{2}- E' + \frac{P'_M - P'}{2} \big] \rangle \nonumber \\
    &&+ \langle (\divprim \bb')\big[R'_H - R_H -\bar \rho \vv \cdot \bb'+ H\big]\rangle +  \langle(\div \bb)\big[R_H - R'_H -\bar \rho \vv' \cdot \bb+ H' \big] \rangle \nonumber \\
    &&+ \frac{1}{2}\langle (\frac{\bb^2}{2} + u) \divprim (\rho' \vv') \rangle + \frac{1}{2}\langle (\frac{\bb'^2}{2} + u') \div (\rho \vv) \rangle - \frac{1}{2}\langle \beta'{}^{-1} \divprim (\rho u' \vv) + \beta^{-1} \div (\rho' u \vv') \rangle \, . \label{iso}
\ea
\end{widetext}

In the limit of $\bb=0$ one can easily recover the exact law of compressible isothermal HD turbulence derived in \cite{galtier_exact_2011}.

\subsubsection{Compressible polytropic MHD turbulence}
In the polytropic case, the state equation is $\gamma P =  c_s{}^2 \rho$, but now the sound speed is no longer constant and depends on density variations, $\gamma$ is the polytropic index. The source terms now are:
\begin{widetext}
\ba
\divl\langle  (1&+&\frac{ \rho' }{\rho}) (P + P_M) \vv' - (1+\frac{ \rho }{\rho'})(P' + P'_M) \vv + \rho' u \vv' - \rho u' \vv \rangle - \langle 2\rho \frac{P'}{\rho'} \divprim \vv' \rangle - \langle 2 \rho' \frac{P}{\rho} \div \vv \rangle \\
=\langle (P_M &+& P - 2 \rho \frac{P'}{\rho'}) \divprim \vv' \rangle + \langle (P'_M + P' - 2 \rho' \frac{P}{\rho}) \div \vv \rangle + \langle (\frac{c_s{}^2}{\gamma} + \frac{\bb^2}{2} + u) \divprim (\rho' \vv') \rangle + \langle (\frac{c'_s{}^2}{\gamma} + \frac{\bb'^2}{2} + u') \div (\rho \vv) \rangle \, . \nonumber
\ea

Consequently, the polytropic MHD exact law reads
\ba
- 2 \varepsilon
    &=&  \frac{1}{2}\divl \langle \big[\delta (\rho \vv ) \cdot \delta \vv  + \delta (\rho \bb ) \cdot \delta \bb  + 2 \delta \rho \delta u \big]\delta \vv - \big[\delta (\rho \bb ) \cdot \delta \vv  + \delta (\rho \vv ) \cdot \delta \bb \big]\delta \bb \rangle \nonumber \\
    &&+ \langle (\divprim \vv')\big[R_E -\frac{R_B + R'_B}{2}- E +  \frac{1}{2}(P_M + P - 2 \rho\frac{c'_s{}^2}{\gamma})\big] +(\div \vv)\big[R'_E -\frac{R'_B + R_B}{2}- E' +  \frac{1}{2} (P'_M + P' - 2 \rho' \frac{c_s{}^2}{\gamma})\big] \rangle \nonumber \\
    &&+\langle (\divprim \bb')\big[R'_H - R_H -\bar \rho \vv \cdot \bb'+ H\big]\rangle +  \langle(\div \bb)\big[R_H - R'_H -\bar \rho \vv' \cdot \bb+ H' \big] \rangle \nonumber \\
    &&+ \frac{1}{2} \langle (\frac{c_s{}^2}{\gamma} + \frac{\bb^2}{2} + u) \divprim (\rho' \vv') \rangle + \frac{1}{2} \langle (\frac{c'_s{}^2}{\gamma} + \frac{\bb'^2}{2} + u') \div (\rho \vv) \rangle\nonumber \\
    &&- \frac{1}{2}\langle \beta'{}^{-1} \divprim (\rho u' \vv) + \beta^{-1}  \div (\rho' u \vv') \rangle \, . \label{polytrope}
\ea
\end{widetext}
Equation (\ref{polytrope}) is the second main result of this work. One can recognize the same structure as in its counterpart describing isothermal CMHD turbulence \citep{andres_alternative_2017}. The first line of the r.h.s. term of equation (\ref{polytrope}) is the usual flux terms that depend only of field increments, the following three lines are the source terms that depend upon the divergence of the Alfv\'en and flow speeds and contain the hybrid term that can be written either as  a source or a flux term, the last line is the $\beta$-dependent term \citep{andres_alternative_2017}. 

Comparing equations (\ref{iso})-(\ref{polytrope}) one can see that the choice of the closure equation (polytropic vs. isothermal) alters the exact law in two locations: in some of source terms that depend on the internal energy or on the polytropic index $\gamma$, and in the internal energy flux term $2 \delta \rho \delta u \delta \vv$ through the dependence of the internal energy on the chosen closure. Quantifying the importance of such dependence on the estimation of the cascade rate in the solar wind is the goal of the next section. 

Finally, we note that in the limit $\bb=0$ one can obtain the following exact law for polytropic HD turbulence:
\begin{widetext}
\ba
- 2 \varepsilon
    &=&  \frac{1}{2}\divl \langle \big[\delta (\rho \vv ) \cdot \delta \vv + 2 \delta \rho \delta u \big]\delta \vv \rangle + \langle (\divprim \vv')\big[R_E - E +  \frac{1}{2}( P - 2 \rho\frac{c'_s{}^2}{\gamma})\big] + (\div \vv)\big[R'_E - E' +  \frac{1}{2} ( P' - 2 \rho' \frac{c_s{}^2}{\gamma})\big] \rangle \nonumber \\
    &&´+ \frac{1}{2} \langle (\frac{c_s{}^2}{\gamma} + u) \divprim (\rho' \vv') + (\frac{c'_s{}^2}{\gamma} + u') \div (\rho \vv) \rangle \, . \label{polytropeHD}
\ea
\end{widetext}

The expression of this law is different from the one derived by \cite{banerjee_kolmogorov-like_2014}, because of the different expressions of the internal energy correlators used in the two derivations ($\rho \sqrt{u} \sqrt{u'} = \frac{\rho C_s C'_s}{\gamma(\gamma-1)}$ instead of $\rho u' = \frac{\rho C'_s C'_s}{\gamma(\gamma-1)}$). The two laws should however be equivalent if tested on numerical simulations as done for IHMHD \citep{ferrand_exact_2019}.

\section{Application to spacecraft observations}\label{obs}

\begin{figure*}[ht]
\centering
\includegraphics[width=\hsize]{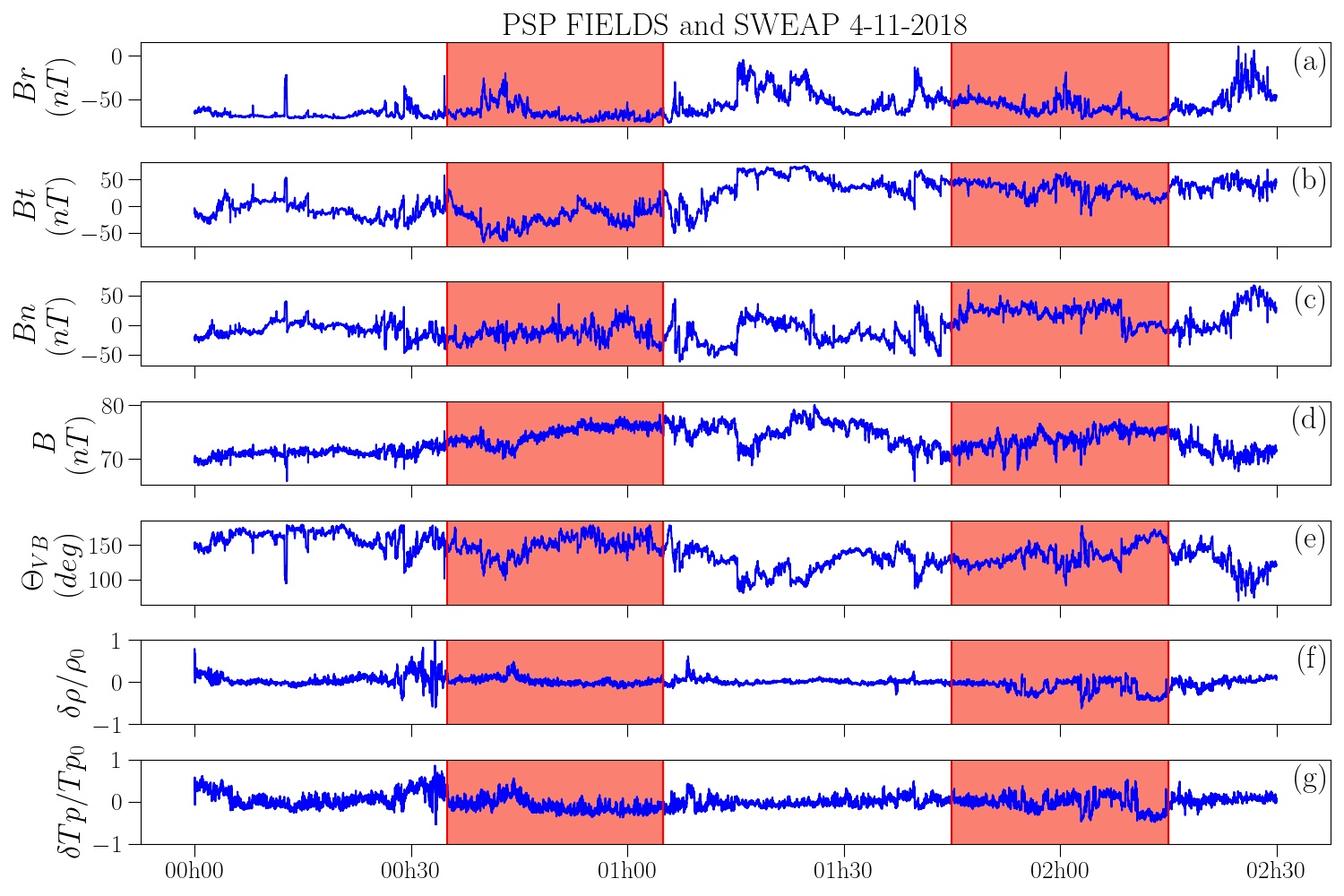}
\caption{The PSP data measured in the inner heliosphere on day 2018-11-04. Panels (a)-(c) are the three magnetic field components (in RTN reference frame), (d) is the corresponding magnitude, (e) is the angle between the fluid velocity and the magnetic field, (f) and (g) are the relative proton density and temperature fluctuations, respectively. The red areas indicate the intervals where the cascade rate is computed. }
\label{fig:panel0}
\end{figure*}

We use Parker Solar Probe (PSP) data recorded on November, 4, 2018 between 0h and 2h30 during its first orbit at around $36$ Rs, Rs is the solar \ADD{radius} \citep{fox_solar_2016}. The data used are those of the magnetic field measured by the FIELDS experiment with a sampling time of $3$ ms \citep{bale_fields_2016} and of the plasma moments (density, ion velocity and temperature) from the SWEAP (Solar Wind Electrons Alphas and Protons) experiment \citep{kasper_solar_2016} sampled at $0.873$ s. In order to compute the cascade rate, which involve correlations between the plasma moments and the magnetic field data we had to re-sample the magnetic field data at the resolution of the SWEAP measurements. The analyzed data are shown in Fig.\ref{fig:panel0}. We analyze two sub-intervals (marked by the red color) that reflect two different levels of density fluctuations: the first one (subset 1) from 0h35 to 1h05 shows very low compressibility (typically $\left <|\delta \rho|/\rho_0\right> \sim 8\%$), while the second one (subset 2) from 1h45 to 2h15 shows larger density fluctuations ($ \left <|\delta\rho|/\rho_0\right> \sim 20\%$). This choice is made to investigate the impact of the different levels of density fluctuations on the cascade rate estimated from isothermal and polytropic CMHD models with respect to that given by the IMHD model. \ADD{The choice of $30$ mn duration guarantees including about three correlation times of the turbulent fluctuations in the analyzed data based on the estimation given by \citet{parashar_measures_2020} (our intervals were included in that study where a correlation time is estimated to $\sim 600s$)}. In Fig.\ref{fig:panel0} we also plot the angle $\Theta_{\bf VB}$ between the local magnetic field and plasma velocity. It is used to check that the two selected periods correspond to relatively stationary angle, a necessary condition to obtain a reliable estimate of the cascade rate as shown in \cite{hadid_energy_2017} (because of the use of the Taylor hypothesis, sampling very different directions with respect to the local magnetic field can result in large fluctuations of the cascade rate).

To estimate the cascade rate we introduce two further assumptions to the model. Considering that the source terms require estimating the divergence of the plasma and the compressible Alfv\'en speeds, which can be done only using multi-spacecraft data \citep{andres_energy_2019}, we assume that the source terms are negligible with respect to the flux terms. This assumption is confirmed in numerical simulations results of sub-sonic MHD turbulence \citep{andres_exact_2018}. Therefore, in the following, we only use the first line in equation (\ref{polytrope}). We further assume the isotropy of the fluctuations, which allows us to integrate in 3D on a ball of radius $\ell$ the reduced form of equation (\ref{polytrope}) and obtain the following expression

\begin{widetext}
\ba
\varepsilon
	 &=& -\frac{3}{4v_0\tau} \left< \left[\delta \left(\rho \mathbf{v}\right) \cdot \delta \mathbf{v}+ \delta \left(\rho \mathbf{v_A}\right) \cdot \delta \mathbf{v_A}\right]\delta {v_\ell} - \left[\delta \left(\rho \mathbf{v_A}\right) \cdot \delta \mathbf{v}+ \delta \left(\rho \mathbf{v}\right) \cdot \delta \mathbf{v_A}\right]\delta {{v_A}}_\ell \right> - \frac{3}{4v_0\tau} \left< 2\delta \rho \delta u \delta {v_\ell} \right> \, 
	 \nonumber \\
	 &=& F_1+F_2  \, , \label{eps_CMHD} 
\ea
\end{widetext}
where $\ell$ denotes the increment direction taken along the mean flow velocity using the Taylor hypothesis $\ell\sim v_0\tau$, $\tau$ being the time lag ($v_0$ is obtained by averaging over each time interval, as all the quantities that are indexed by $0$). The difference between the isothermal and the polytropic models hides in the form of the specific internal energy $u$ given by $u = c_s^2 \ln (\rho / \rho_0)$ in the isothermal case and $u = (c_s^2 - c_{s0}^2) / (\gamma(\gamma-1))$ in the polytropic one (with $\gamma\neq1$). This choice is made to ensure a consistency between the two models regarding the initial (background) internal energy ($\rho=\rho_0 \rightarrow u=0$). The sound velocity $c_s$ comes from the perfect gas equation, $c_s^2 = \gamma k_B T_p / m_p$ with $\gamma = 1$ for the isothermal case and $\gamma = 5/3$ for the polytropic, $T_p$ and $m_p$ the proton temperature and mass. In equation (\ref{eps_CMHD}) the term $F_1$ refers to the modified (i.e., compressible) Yaglom term given by equation (\ref{imhd}), while $F_2$ refers to the internal energy one, which has no counterpart in the IMHD theory. Within the reduced equation (\ref{eps_CMHD}) the difference in estimating the polytropic and isothermal cascade rate would come only from the term $F_2$.

\begin{figure*}[ht]
\centering
\includegraphics[width=\hsize]{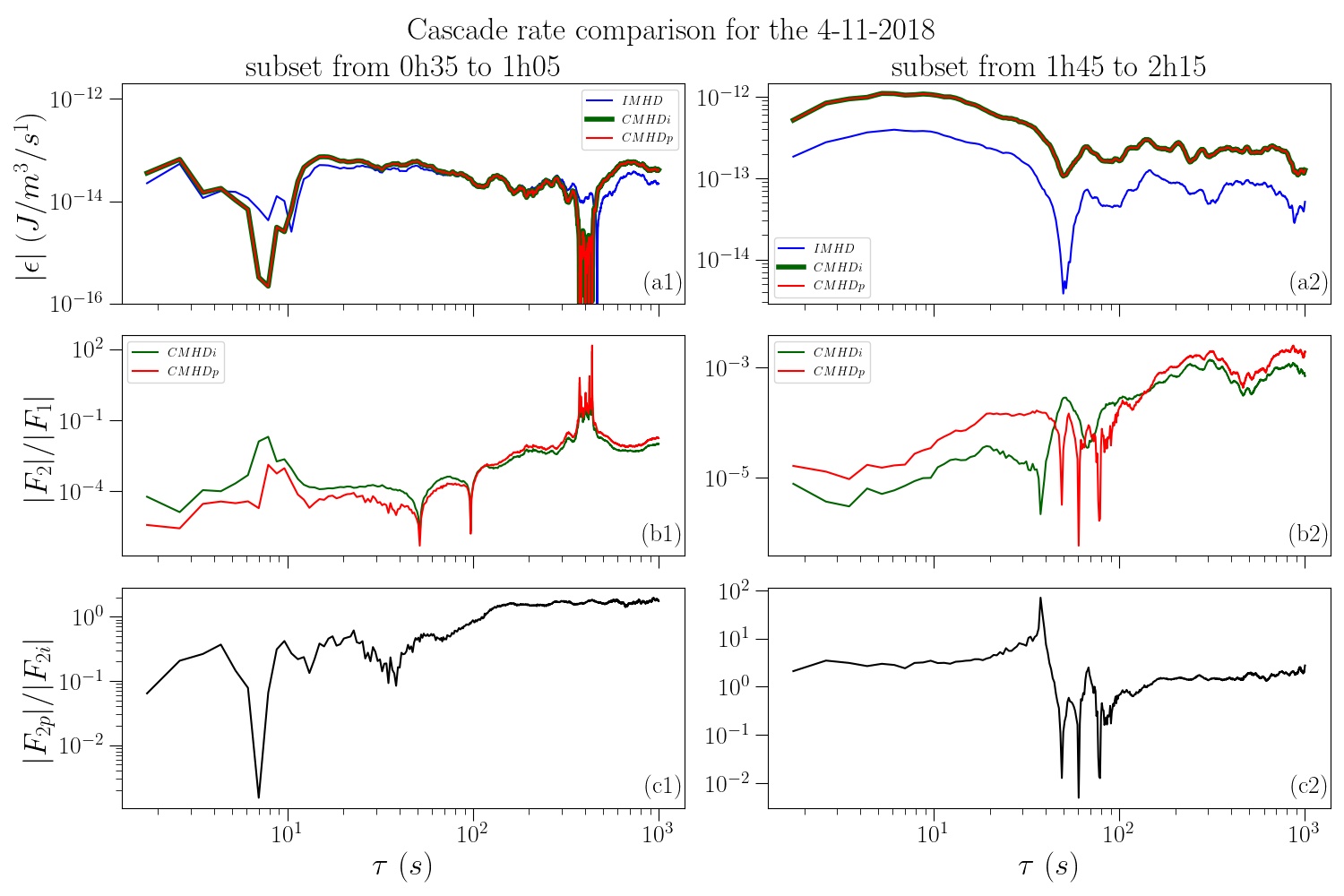}
\caption{Comparison of the computed cascade rate for subset 1 (0h35-1h05, left) and subset 2 (1h45-2h15, right). Panels (a1)-(a2) represent the absolute value of the cascade rate given by each model (incompressible in blue, isothermal in green and polytropic in red), (b1)-(b2) are the ratios between the internal energy term $F_2$ and the compressible Yaglom term $F_1$ for isothermal (green) and polytrope (red) description,  (c1)-(c2) are the ratios between the polytropic  ($F_{2p}$) and isothermal ($F_{2i}$) estimations of the internal energy term.}
\label{fig:result_subset}
\end{figure*}

Figure \ref{fig:result_subset} shows the results obtained from the two subsets of data on the estimation of the full cascade rate using the IMHD, polytropic CHMHD and isothermal CHMHD, and on the separate contribution of each term, $F_1$ and $F_2$. \ADD{Since we are primarily interested in the possible impact of the equation of state on the cascade rate, this first application of the general exact law to real data is limited to the evaluation of the absolute values of the cascade rate. Signed cascade rate and the subsequent question of the forward vs. inverse cascade, which requires a much larger data samples to ensure statistical convergence \citep{coburn_third-moment_2015,hadid_energy_2017}, is thus not treated here. In the first data set (Fig. \ref{fig:result_subset}(a1)) the cascade rate changes sign at $\tau\sim 400$ sec and $\tau\sim 10$ but remains nearly constant between the two time lags, which would define the inertial range where the exact law can be evaluated. The large scale sign change is due to a brief variation in the Yaglom term $F_1$ that appears by an increase Fig. \ref{fig:result_subset}(b1) and a decrease of the incompressible and compressible results Fig. \ref{fig:result_subset}(a1). In the second data set (Fig. \ref{fig:result_subset}(a2)) the cascade rate shows no sign change, but the inertial range would be limited to the range $\tau\sim800-50$ sec.}

Figure \ref{fig:result_subset}(a2) emphasizes the role of density fluctuations in amplifying the compressible cascade rate by a factor $\sim 3$ with respect to the incompressible one. This contrasts with the data in subset 1 (Fig. \ref{fig:result_subset}(a1)) where the compressible and incompressible cascade rate coincide, in agreement with the lower density fluctuations observed during this time interval ($\left <|\delta \rho|/\rho_0\right> \sim 8\%$). This result confirms the previous findings in the solar wind \citep{banerjee_scaling_2016,hadid_energy_2017,andres_evolution_2021}. Figs. \ref{fig:result_subset}(a1)-\ref{fig:result_subset}(a2) show further that the two compressible models (polytropic and isothermal) give essentially the same cascade rate. The reason for the convergence of the two models can be seen in Figs. \ref{fig:result_subset}(b1)-\ref{fig:result_subset}(b2): the contribution of the internal energy term $F_2$ in both subsets is negligible with respect to the compressible Yaglom term $F_1$ \ADD{(note that the burst of the ratio $F_2/F_1$ observed at $\tau \sim 400$ sec is not caused by an increase of the internal energy term but rather by the decline of the Yaglom term $F_1$ because of its sign change at the same time lag mentioned above)}. From this observation one can conclude that the amplification in the total cascade rate seen in Fig. \ref{fig:result_subset}(a2) comes from the contribution of the density fluctuations to the Yaglom term $F_1$. However, even if in these two particular cases the isothermal and polytropic closure yield the same estimation of the cascade rate (since $F_2<<F_1$), one can still examine the impact of each closure on the internal energy term $F_2$. The result is given in Figs. \ref{fig:result_subset}(c1)-\ref{fig:result_subset}(c2). We observe that at the largest scales ($\tau \gtrsim 100 $ sec) the two closures provide nearly the same estimation of the internal energy term $F_2$. However, at smaller scales there is about an order of magnitude difference in the estimates given by the two models: in subset 1 the polytropic closure provides a lower contribution to the cascade rate than the isothermal, while the opposite trend is observed in subset 2.
However, a firm conclusion cannot be reached about the physical origin of the difference between the two models, which would require analyzing a larger data sample. Nevertheless, this first application of our new model to spacecraft observations shows that the choice of the ``thermodynamics'' can impact the estimation of the cascade rate in real data. This impact is likely to be higher in the magnetosheath where density fluctuations are generally larger than those reported here, in which case both $F_1$ and $F_2$ would play a leading role in the turbulent cascade \citep{andres_energy_2019}. 

\section{Discussion and conclusions}\label{discussion}
In this work we provided a general theoretical framework to derive exact relations for homogeneous compressible turbulent flows described within the MHD model. This general framework relies only on the assumption of constant entropy (in the inertial range), which allowed us to introduce a general equation for the internal used in the derivation, which encompasses the classical isothermal and polytropic closures (in addition to the incompressible limit). We showed that this formalism allows, when the closure equation in stated in the exact relations, recovering all known limits of incompressible and compressible isothermal MHD laws, and to obtain a new one that describes polytropic MHD turbulence. We found that the choice of the state equation impacts both the flux (through the internal energy) and the source terms. To quantify such an impact we applied a reduced form of the two theoretical models (where only flux terms were retained) to \ADD{two intervals of PSP observations taken in the inner heliosphere}. We showed that while the overall impact of the state equation on the cascade rate is negligible (because of the relatively low density fluctuations in the solar wind), it does however influence the internal energy term. \ADD{The same conclusion is obtained from a few intervals of MMS data measured in the magnetosheath (not shown) where density fluctuations were higher (up to $50\%$), but based only on the estimation of the flux terms. Nevertheless, a firm conclusion as to how the two compressible models would impact the cascade rate estimates in spacecraft observations requires analyzing a larger data sample. The use of multi-spacecraft MMS data should allow us to obtain a full answer to this question, through the estimation of both the flux {\it and} source terms.} 

\acknowledgments{P. Simon is funded through a DIM ACAV+ PhD grant. The authors acknowledge the Johns Hopkins Applied Physics Laboratory for designing, building, and now operating Parker Solar Probe as part of NASA’s Living with a Star (LWS) program (contract  NNN06AA01C). They also acknowledge the use of data from FIELDS (\url{http://research.ssl.berkeley.edu/data/psp/data/sci/fields/l2/}) and SWEAP (\url{http://sweap.cfa.harvard.edu/pub/data/sci/sweap/}) instruments.}

\appendix
\begin{widetext}
\section{Derivation of the equation of kinetic energy correlator}
\label{ann:details}
The exact law given in the main text is obtained from the equation governing the temporal evolution of the total correlator $2 R_{tot} = \langle  (R_k + R'_k) + (R_B + R'_B) + (R_u + R'_u) \rangle/2$ (i.e., Eq. (\ref{cor})). The strategy for deriving such an equation is the same for the three terms related to the kinetic, magnetic and internal energy correlators. Therefore, here we provide only the details of the derivation of the kinetic energy correlator $(R_k + R'_k)$.
Using equations (\ref{mhd1}), (\ref{mhd3}) and (\ref{ther5}), the momentum equation (\ref{mhd2}) can be written under the following form :
\ba
\partial_t \vv + \div (\vv \vv) - \vv \div \vv &=& \div (\bb \bb) - 2 \bb \div \bb - \grad (\frac{P + P_M}{\rho})   - (1 + \frac{P_M}{P}) \grad u + \frac{{\bf d_k}}{\rho} + \frac{{\bf f}}{\rho} \, . \label{ann:mhd2'}
\ea

Writing (\ref{mhd2}) at the point $\textbf{x}$ multiplied by $\vv'$ then at the point $\textbf{x}'$ multiplied by $\vv$, and (\ref{ann:mhd2'}) at the point $\textbf{x}$ multiplied by $\rho'\vv'$ then at the point $\textbf{x}'$ multiplied by $\rho \vv$, summing the four terms and using the property $\partial_t \langle \rangle = \langle \partial_t  \rangle$ yield the following expression for the time evolution of the kinetic energy correlator
\ba
2 \langle \partial_t (R_k + R'_k) \rangle &=& \langle \partial_t (\rho \vv \cdot \vv' + \rho' \vv' \cdot \vv ) \rangle  \nonumber \\ 
&=& \langle \partial_t (\rho \vv) \cdot \vv' + \rho \vv \cdot \partial_t \vv' + \partial_t (\rho' \vv') \cdot \vv + \rho' \vv' \cdot \partial_t \vv \rangle  \nonumber \\ 
&=& \langle (-\div (\rho \vv \vv) + \div (\rho \bb \bb) - \grad (P + P_M) + {\bf d_k} + {\bf f}) \cdot \vv' \rangle  \nonumber \\ &&+ \langle \rho \vv \cdot ( - \divprim (\vv' \vv') + \vv' \divprim \vv' + \divprim (\bb' \bb') - 2 \bb' \divprim \bb' - \grad' (\frac{P' + P'_M}{\rho'})   - (1 + \frac{P'_M}{P'}) \grad' u' + \frac{{\bf d'_k}}{\rho'} + \frac{{\bf f'}}{\rho'}) \rangle  \nonumber \\ &&+ \langle (-\divprim (\rho' \vv' \vv') + \divprim (\rho' \bb' \bb') - \grad' (P' + P'_M) + {\bf d'_k} + {\bf f'}) \cdot \vv \rangle  \nonumber \\ &&+ \langle \rho' \vv' \cdot (- \div (\vv \vv) + \vv \div \vv + \div (\bb \bb) - 2 \bb \div \bb - \grad (\frac{P + P_M}{\rho})   - (1 + \frac{P_M}{P}) \grad u + \frac{{\bf d_k}}{\rho} + \frac{{\bf f}}{\rho} ) \rangle \, . \nonumber
\ea
Using the definition for any entity A: $\partial_x A' = \partial_{x'} A = 0$ one gets :
\ba
2\langle \partial_t (R_k + R'_k) \rangle &=& \langle -\div (\rho \vv \cdot \vv' \vv) + \div (\rho \bb \cdot \vv' \bb) - \div ((P + P_M)\vv') + {\bf d_k}\cdot \vv' + {\bf f} \cdot \vv' \rangle  \nonumber \\ 
    &&+ \langle - \divprim (\rho \vv \cdot \vv' \vv') + \rho \vv \cdot \vv' \divprim \vv' + \divprim ( \rho \vv \cdot \bb' \bb') - 2  \rho \vv \cdot \bb' \divprim \bb' - \divprim (\frac{P' + P'_M}{\rho'} \rho \vv)  \nonumber \\ 
    && \qquad \qquad  - (1 + \frac{P'_M}{P'}) \divprim  (\rho u' \vv)  +  \rho \vv \cdot \frac{{\bf d'_k}}{\rho'} +  \rho \vv \cdot \frac{{\bf f'}}{\rho'} \rangle  \nonumber \\ 
    &&+ \langle -\divprim (\rho' \vv' \cdot \vv  \vv') + \divprim (\rho' \bb' \cdot \vv  \bb') - \divprim ((P' + P'_M)\cdot \vv ) + {\bf d'_k}\cdot \vv  + {\bf f'}\cdot \vv \rangle  \nonumber \\ 
    &&+ \langle - \div (\rho' \vv' \cdot \vv \vv) + \rho' \vv' \cdot \vv \div \vv + \div (\rho' \vv' \cdot \bb \bb) - 2 \rho' \vv' \cdot \bb \div \bb - \div (\frac{P + P_M}{\rho}\rho' \vv')   \nonumber \\ 
    && \qquad \qquad  - (1 + \frac{P_M}{P}) \div (\rho' u \vv') + \rho' \vv' \cdot \frac{{\bf d_k}}{\rho} + \rho' \vv' \cdot \frac{{\bf f}}{\rho}  \rangle \, . \nonumber
\ea

Thanks to homogeneity assumption: $\langle \div \rangle = - \divl \langle \rangle$ and $\langle \divprim \rangle = \divl \langle \rangle$ and after rearranging some terms, we obtain the following equation :
\ba
2\langle \partial_t (R_k + R'_k) \rangle
    &=& \divl \langle \rho \vv \cdot \vv' \vv - \rho \vv \cdot \vv' \vv' - \rho' \vv' \cdot \vv \vv' + \rho' \vv' \cdot \vv \vv \rangle \nonumber \\
    &&- \divl \langle \rho \bb \cdot \vv' \bb - \rho \vv \cdot \bb' \bb' - \rho' \bb' \cdot \vv \bb' + \rho' \vv' \cdot \bb \bb \rangle \nonumber \\ 
    &&+ \divl \langle (P + P_M)\vv' - (P' + P'_M)\vv - \frac{\rho}{\rho'} (P' + P'_M) \vv +  \frac{\rho'}{\rho} (P + P_M) \vv' - \rho u' \vv +  \rho' u \vv' \rangle \nonumber \\
    &&+ \langle \rho' \vv' \cdot \vv \div \vv + \rho \vv \cdot \vv' \divprim \vv' - 2\rho' \vv' \cdot \bb \div \bb - 2\rho \vv \cdot \bb' \divprim \bb' - \frac{P'_M}{P'} \divprim (\rho u' \vv) - \frac{P_M}{P} \div (\rho' u \vv') \rangle \nonumber \\
     &&+ \langle {\bf d_k} \cdot \vv' + \frac{\rho}{\rho'}{\bf d'_k} \cdot \vv + {\bf d'_k} \cdot \vv +\frac{\rho'}{\rho}{\bf d_k} \cdot \vv' + {\bf f} \cdot \vv' + \frac{\rho}{\rho'}{\bf f'} \cdot \vv + {\bf f'} \cdot \vv + \frac{\rho'}{\rho}{\bf f} \cdot \vv' \rangle \, , \label{ann:eq-corK}
\ea

which is identical to equation (\ref{eq-corK}) in the main text. In the first line of the equation (\ref{ann:eq-corK}), we recognize the flux terms of the developed form of the divergence of the structure function $\langle \delta(\rho \vv ) \cdot \delta \vv \delta \vv \rangle$: 
\ba
\divl \langle \delta(\rho \vv ) \cdot \delta \vv \delta \vv \rangle &=& \divl \langle \rho \vv \cdot \vv' \vv - \rho \vv \cdot \vv' \vv' - \rho' \vv' \cdot \vv \vv' + \rho' \vv' \cdot \vv \vv \rangle + \langle \rho \vv \cdot \vv \divprim \vv' + \rho' \vv' \cdot \vv' \div \vv \rangle \, , \label{ann:fctstruct}
\ea
where by homogeneity, $\divl \langle \rho' \vv' \cdot \vv' \vv' - \rho \vv \cdot \vv \vv \rangle = 0$ and $\divl \langle \rho \vv \cdot \vv \vv' - \rho \vv \cdot \vv \vv' \rangle = \langle \rho \vv \cdot \vv \divprim \vv' + \rho' \vv' \cdot \vv' \div \vv \rangle $. 

Then the equation (\ref{ann:eq-corK}) becomes : 
\ba
2\langle \partial_t (R_k + R'_k) \rangle
    &=& \divl \langle \delta(\rho \vv ) \cdot \delta \vv \delta \vv  \rangle \nonumber \\
    &&- \divl \langle \rho \bb \cdot \vv' \bb - \rho \vv \cdot \bb' \bb' - \rho' \bb' \cdot \vv \bb' + \rho' \vv' \cdot \bb \bb \rangle \nonumber \\ 
    &&+ \divl \langle (P + P_M)\vv' - (P' + P'_M)\vv - \frac{\rho}{\rho'} (P' + P'_M) \vv +  \frac{\rho'}{\rho} (P + P_M) \vv' - \rho u' \vv +  \rho' u \vv' \rangle \nonumber \\
    &&+ \langle - \rho' \vv' \cdot \delta \vv \div \vv + \rho \vv \cdot \delta \vv \divprim \vv' - 2\rho' \vv' \cdot \bb \div \bb - 2\rho \vv \cdot \bb' \divprim \bb' - \frac{P'_M}{P'} \divprim (\rho u' \vv) - \frac{P_M}{P} \div (\rho' u \vv') \rangle \nonumber \\
     &&+ \langle {\bf d_k} \cdot \vv' + \frac{\rho}{\rho'}{\bf d'_k} \cdot \vv + {\bf d'_k} \cdot \vv +\frac{\rho'}{\rho}{\bf d_k} \cdot \vv' + {\bf f} \cdot \vv' + \frac{\rho}{\rho'}{\bf f'} \cdot \vv + {\bf f'} \cdot \vv + \frac{\rho'}{\rho}{\bf f} \cdot \vv' \rangle \, . \label{ann:eq-corK2}
\ea

The three first lines of Eq. (\ref{ann:eq-corK2}) are expressed as a divergence of flux terms, while the fourth line as source terms (i.e., proportional to field divergence such as $\div \vv$ or $\div \bb$). The last line contains the forcing and dissipation terms. Note however that this choice of presentation is not unique as some of the flux terms above can be expressed as sources (and vice-versa), hence the terminology of hybrid terms introduced by \cite{andres_alternative_2017}.

\end{widetext}

\bibliography{Ref}
\end{document}